# Current and field stimulated motion of domain wall in narrow permalloy stripe

L.S. Uspenskaya, S.V. Egorov

*Institute of Solid State Physics Russian Academy of Sciences,142432, Chernogolovka, Moscow region, Russia*

Abstract— Of the new types of cryoelectronic devices under development, including phase shifters, giant magneto-resistance switches, diodes, transistors, and memory cells, some are based on hybrid superconductor-normal metal or superconductor-ferromagnet films. Control of these devices is realized by means of pulses of voltage, light, or magnetic field. Spin-polarized current may be used to switch low-temperature devices, as in spin-electronic devices. In the superconducting layer, the current is dissipation less, which would bring large reduction of energy consumption. We demonstrate that magnetic domain walls in bilayer niobium-permalloy stripes are shifted by electrical current along the stripe even at low temperature, with the niobium in the superconducting state. The wall motion in response to current pulses is quite different from that induced by a magnetic field pulses only. The effect could be used to create a new type of sequentially switched serial devices because of very high value of the wall velocity, which excides by many orders of magnitude the velocity of the wall moved with magnetic field pulses.

## Introduction

Switching of the magnetic state of the ferromagnet nanostructures by spin-polarized current is promising for applications in spintronics [1, 2, 3, 4]. It allows the developing of new efficient elements for data recording and readout, novel types of logic devices and switches [5, 6, 7]. Therefore, the influence of the spin polarized current on the magnetic domain structure as well as current driven motion of magnetic domain walls (DWs) is studied extensively. Different mechanisms underlying the action of the electric current on the DWs are determined [8, 9, 10, 11, 12, 13]; the velocity of current induced motion is measured [13]; the velocity restriction due to oscillatory behavior is shown [14].

Phase shifters, giant resistance switches, diodes, transistors, and memory cells based on hybrid superconductor-ferromagnet structures [15, 16, 17] are developed in cryoelectronics; all these devices are switched by magnetic field. However, it looks very attractive to apply the spin-polarized current instead of the magnetic field, first of all, because of dissipation less character of the current flow in the superconductors and because of simplification of architecture of control circuits.

In our previous papers [18, 19] we have shown that magnetic domain wall in hybrid permalloy-niobium structure is shifted with relatively low current pulses, $j_c \sim 10^7$ A/cm$^2$. Extremely high wall velocity, $V_{max} \sim 4000$ m/s, was reached at temperature near 6 K. Moving with such velocity, the wall widens up to hundreds of micrometers and turns into domain with transverse to the permalloy stripe magnetization [19]. The transformation resembles one observed in [20]. However, the current strength there was $10^{12}$ A/m$^2$ and produced the field higher than coercivity for field driven domain wall motion, while in our case the transformation was observed with the current pulses about $j = 2 \times 10^7$ A/cm$^2$, which produce the field by a factor of ten smaller than coercivity field. Differently from [20], the wall transformation observed in our case was reversible: New formed domain shrunk with opposite current pulses. The wall velocity decreased twenty times with temperature increase from 6 K up to 300 K according to power law despite of the decrease of coercivity field and growth of the velocity of field stimulated wall motion [18]. We could not explain obtained results in frame of existing theories, but the observed phenomenon looked very attractive as it opened possibility to develop a new type of sequentially switched serial devices, the advantage of which would be provided by high rate of switching due to high wall velocity, which exceeds by many orders of magnitude the velocity of the wall moved with magnetic field pulses.

In this work we report our new experimental results on comparison of pulse excitation of magnetic domain walls with electric current and magnetic fields in bilayer permalloy-niobium tapes. Using new type of structure, we demonstrate that discussed above transformation of the wall with current pulses still takes place. We measure the temperature variations of the coercivity, of the delay in the DW motion following pulse excitation, and of the velocity of leading and tailoring part of the domain wall. We show that the velocity of both parts of the DW driven by the current is much higher than the velocity of the wall shifted by the magnetic field. We show that the transformation of current driven DW is remarkably different from those induced by the magnetic field of any direction. Finally, we suggest the way to use the observed effects in technical devices.

### Experiments

The experiments are performed on the bilayer niobium-permalloy stripe with the width of about 5 μm and the thicknesses of permalloy upper layer and niobium sublayer of about 40 nm and 100 nm, respectively, figure 1. The length of the stripe is about 1 mm. The structure is fabricated by the lift-off lithography following magnetron evaporation onto oxidized silicon substrate at room temperature. First, the contact structure is fabricated (figure 1b). After that the bilayer bridge shaped as the narrow stripe (the sample under the study) is fabricated in a single vacuum cycle in the presence of 600 Oe in-plane magnetic field between the contacts, figures 1a, 1b. Such procedure provides the in-plane orientation of spontaneous magnetization of permalloy layer and its parallelism to the stripe.

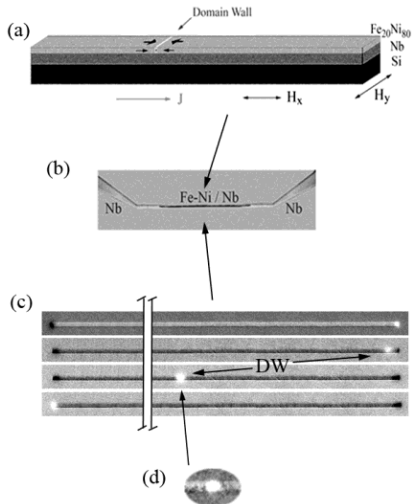

Fig. 1. (a) – Sketch of the sample – bilayer rectangular stripe with dimensions about (40+100) nm × 5 μm × 1 mm; (b) - real image of the studied structure, (c) – the images of the magnetic pattern taken in zero field, under the applied field about 0.80 × 10³ A/m, 0.80 × 10³ A/m, and 0.81 × 10³ A/m (some part of the image is cut off to make visible both ends of the sample), and (d) - zoomed image of the wall. The DW is seen as white sport shifted from the right end of the sample to the left. Black and white contrast at the sample ends correspond to oppositely directed stray field.

The visualization of the magnetic domain structure and domain wall motion is performed by means of yttrium-iron garnet film (indicator) placed on the top surface of the sample [18,19] and by bitter decoration technique [21]. Optical observations are performed in polarized light microscope. The technique visualizes the stray fields both above the domain walls and at the edges of the sample (figure 1c). These fields give rise to the local rotation of the polarization plane for the light reflected from the indicator. The rotation leads to the local brightening or darkening of the image obtained in the reflected polarized light with the polarizer and analyzer uncrossed by five degrees, figures 1c, 2a, and 2b. The resolution of the method is restricted principally by an optical one, i.e. by ~ 0.6 μm. Real resolution is even worse because there is always some unavoidable distance between the indicator and the sample giving a magnetic gap. Besides, the saturation field of the yttrium-iron garnet film is about 8 × 10⁴ A/m. The saturation magnetization of permalloy is about 1 T. Therefore, the stray field near the sample edges or around the 180° domain walls saturates the magnetization of the garnet film in the area, which is much wider than the width of DWs. Thus, obtained images often do not reflect the real width of the DW but rather point out the wall position and the symmetry of magnetization distribution around the wall, figures 1c, 2a, and 2b. The Bitter method is free from these limitations. It could reproduce fine features of the domain structure with much better resolution if one uses nanoparticles [21]. We have applied the 10% water solution of commercial ferrofluid EMG 705 with 10 nm nanoparticles having spontaneous magnetization about 22 mT. Comparing the magnetooptic observations with Bitter patterns, we ascertained the type of magnetic domain walls and the view of closure-domains of our narrow stripe-like samples. Unfortunately, Bitter method could not be applied to study dynamic processes.

Both static and dynamic magnetooptic images were recorded in real time using an SDU285 digital camera with exposition from 10 ms down to 1 μs. The registration was performed in the moments of time given by synchronization signal. To register shorter events, we applied illumination by pulsed laser beam. Digital subtraction of background was performed to increase the contrast of the image.

The DWs velocity was estimated from the measurements of their shift between two serial video-frames trapped synchronously with the leading and trailing edges of exciting pulses, like it was done in Ref. [19]. Other features of the visualization technique and the methods of determining the type of domain walls applied here are described in details in [22].

### Results and discussion

Initially, the stripes are in single domain state with magnetization aligned with the stripe length (figure 1c, first image). Such magnetization alignment is in agreement with sample fabrication in the presence of the magnetic field, which provides the in-plane anisotropy about 160 A/m. The alignment with the stripe length is also supported by large aspect ratio of the stripe, about 1 : 125 : 25000. So, the single domain state is equilibrium one, and the stripe returns to this state following quasi-static magnetization reversal by the field applied in any direction. The freezing of some structure with domain walls is possible only by short enough pulsed magnetic field, figure 2.

Domain walls are formed at both stripe edges. Further they shift into the stripe following pulse magnetic field applied along the stripe $H_x$. The wall formation occurs with some delay relative to the beginning of the pulse, figure 3a. This delay decreases with an increase of the field and with rise of temperature $T$, but it always exceeds $10^{-4}$ s if the field is below $8 \times 10^3$ A/m (larger field causes multiple nucleation of the walls producing the pattern like boiling water). As usual, the wall motion induced by the field pulse is not uniform. After some delay, the wall velocity reaches the maximum and then it slows down. The dependences of the maximum velocity upon the field strength $V_H(H)$ taken in the temperature range from 300 K down to 12 K are shown in figure 3 b; similar dependence was obtained in [18] on wider stripe. Temperature dependences of the maximum velocity $V_H(T)$ and coercivity $H_c(T)$ of DWs are presented in figure 4. They practically coincide with those obtained on meander-shaped permalloy-niobium structure with line width of 7 μm studied earlier [18]. The highest velocity also does not exceed 20 m/s; this value is reached at $T = 300$ K. The dependence $V_H(T)$ obeys the power law $\sim T^{1.3}$. The coercivity exhibits an exponential growth with cooling; it ranges from 100 A/m at 300 K up to 4500 A/m at 12 K (similar behavior was reported in [23, 19]).

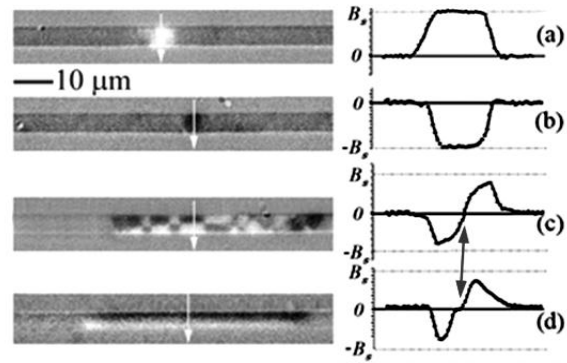

Fig. 2. Magnetic domain patterns (left panels) formed in the permalloy strip following (a, b) the magnetic field pulse applied along the permalloy stripe ($H_x$), (c) perpendicular to the stripe ($H_y$), and (d) the current pulse flowing along the stripe. Intensity profiles of the stripe images taken in the directions shown by the white arrows are presented to the right of the corresponding images. The intensity is proportional to the square of the induction. Note that the induction $B_s$ in the single domain walls (a,b) is larger than in the domains (c) and in the stray fields near the sample (d). The images are taken at 13 K.

Applying the magnetic field along the stripe, we always obtain the DWs of the same type which is the Bloch type, we believe. There are two arguments supporting this statement. First, the image of the static DW looks like symmetric bright spot, figure 1d. There are no traces of the perpendicular to the stripe in-plane component of magnetization. Such image could be obtained just in case of large out-of-plane component of magnetization in the wall, i.e. the wall is not of Neel type. Bitter technique shows the wall image as the straight chain of nanoparticles arranged perpendicular to the stripe, i.e. the wall is not a vortex or a set of vortices like was observed in 1 μm wide permalloy wire grown on silicon substrate in Ref. [20] and like it is predicted by simulations using OMMMF program. Second, we analyzed previously [22] the types of DWs in permalloy nanofilms fabricated by magnetron spattering on silicon substrate and niobium substrates of different thicknesses. We came to the conclusion that the Bloch type walls are nucleated in films and wide stripes in permalloy layer fabricated on 100 nm thick niobium sublayer because of the roughness of the niobium, which grows proportionally to the thickness of the niobium due to its column structure. Thus, the origin of the difference in the domain wall structure in our samples and in [20] is in because of different substrates on which permalloy is grown. So, we have the straight domain walls carrying perpendicular magnetizationin in our samples.

If we apply the in-plane magnetic field at some angle to the stripe (different from 90°), the magnetization reversal also proceeds via the nucleation of the domain walls at the stripe edges and their motion. The coercivity grows with the angle increase, figure 5a, but it is not proportional to the calculated field projection to the stripe direction (dashed lines in figure 5a and figure 5b). Thus, the perpendicular component of the magnetic field modifies the wall properties, e.g. because of the change in the DW structure. The deviation of the in-plane field toward the stripe normal reduces the coercivity nearly twice. Besides, it should increase the wall velocity. As shown in [24], the wall velocity rises from 300 m/s up to 900 m/s under the perpendicular field of about 6 kA/m in 10 nm thick and 300 nm wide permalloy tape.

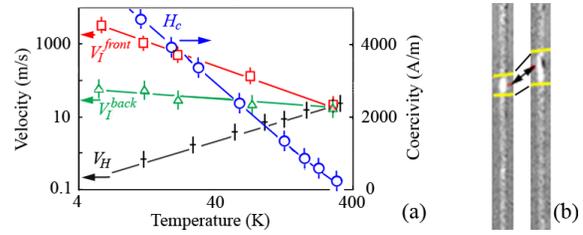

Fig. 4. (a) -Temperature dependences of the coercivity $H_c$ (circles), maximum domain wall velocity moving under the magnetic field $V_H$ (crosses), and the velocity of leading edge of the wall $V_I^{front}$ (squares) and trailing edge of the wall $V_I^{back}$ (triangles) moving under the current pulses, and (b) – two frames showing the widening and the difference in the shift of leading and tailoring parts of the wall moving with current pulse j = 2 × 10$^7$ A/cm$^2$. On the left – the images of the stripe with moving domain wall, before the current pulse and just after it. The lines between images are drawn to show the shift of leading and tailoring parts of the wall

When we apply the in-plane field exactly perpendicular to the stripe ($H_y$), the residual domain pattern depends upon the field strength and pulse duration. The single domain state is reached after the quasi-static process. Application of the short pulses of the field with the strength above 24000 A/m, results in the frozen domain pattern. One example of residual domain structure is given in figure 2c. One can see the microdomains with out-of-plane magnetization vector. The "anti-symmetry" line is located near the center line of the stripe sample. The magnetization vector rotates by 180°, which is evident from the corresponding induction profile, figure 2c. The structure could be equally treated as a set of either microdomains or magnetic vortices, which enter the stripe from the long edges as a result of the demagnetizing field. The latter one is inhomogeneous near the edges due to roughness inevitably arising during the lift-off fabrication. Note, we cannot "open" existing domain wall by any magnetic field. We can either move it along the stripe or destroy it by strong enough perpendicular field.

The applied pulses of electric current lead to the displacement of the domain walls if the current density exceeds $j_0 = 10^7$ A/cm$^2$. The direction of the displacement is determined by the direction of the current. Differently from field excitation, we do not manage to record any delays in the starting of DWs motion after applying the electric current pulses, i.e. the delay does not exceed $10^{-6}$ s, which is much less than the delay at the magnetic field excitation. The velocity of the current stimulated motion is of the order of the velocity of field driven motion at 300 K. It increases rapidly with cooling following the power law $\sim T^{1.4}$. The highest velocity is as large as 4000 m/s at 6 K, figure 3. However, this value concerns the velocity of the leading edge of the wall $V_I^{front}$ only. The trailing edge of the wall moves much slower than the leading edge. Its velocity does not exceed 50 m/s, but it still remarkably exceeds the velocity of the wall shifted with field pulses.

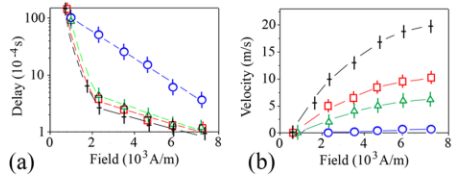

Fig. 3. Field strength dependence of the (a) delay of the DW motion start relative to the beginning of the field pulse (pulse leading-edge time and pulse length are about $10^{-6}$ s and $10^{-2}$ s, respectively) and (b) the DW velocity. The data are measured at temperature of 300 K (black crosses), 90 K (red squares), 60 K (green triangles), and 12 K (blue circles).

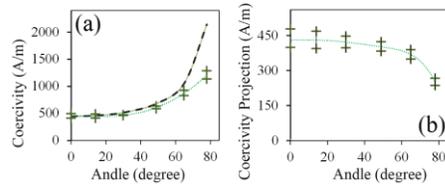

Fig. 5. Influence of the in-plane field direction on the magnitude of coercivity $H_c$. (a) Angular dependence of the coercivity $H_c$. Experimentally determined values are shown by the crosses and the calculated ones – by dashed line. The angle is measured relative to $H_x$ direction (see figure 1b). (b) Angular dependence of the projection of the coercive field on the stripe direction. There should not be angular dependence, if the perpendicular component of the field does not affect the domain wall motion

The difference in the velocities of the leading and tailing parts causes the extension of the DW and formation of a new domain between leading and tailing parts of the wall, figure 2d. The magnetization in this new domain is located in the stripe-plane perpendicular to its length ($M_y$), i.e. perpendicular to the principal direction of magnetization. Such orientation of the magnetization vector should not be stable because of large value of the aspect ratio. Nevertheless, the domain survives till the inverse current pulse or saturated magnetic field is applied. Note, the domains of this type are never observed in our narrow permalloy stripes during or after magnetization by the magnetic field. So, this is a specific feature of the response to the current pulses.

Small pulses of the inverse current shorten the domain. Large pulses collapse it. The length of the domain is controlled by the parameters of the direct and inverse current. The length could reach up to 0.5 mm at 6 K under the current pulse of the magnitude about $j_0 = 5 \times 10^7$ A/cm$^2$ with the duration of 100 ns, figure 2d.

One could suppose that the origin of the domain formation is the in-plane magnetic field produced by the current. Really, the current flows in both permalloy and niobium layers. Therefore it produces the field $H_y$ in the permalloy layer having the strength about 30 kA/m with the current about $j_0 = 5 \times 10^7$ A/cm$^2$. We have mentioned above that such field itself (without electric current flow) causes just reversible rotation of magnetization. Much large field forms the domain pattern like one shown in figure 2c, which is principally different from the pattern in figure 2d: There are domains with perpendicular magnetization in figure 2c and one domain with in-plane magnetization in figure 2d. Therefore, the field cannot be the reason of domain formation, but its existence could influence the velocity between the leading and tailoring parts of the wall as this field favors the widening of the area with perpendicular to the stripe in-plane magnetization and acts into opposite directions on the leading and tailoring parts of the inclined wall respectively accelerating and decelerating them.

The origin of the huge increase in the velocity of the DW driven by electrical current at low temperature still remains an open question. The attempt to explain it via increase in the density of spin-polarized electrons is not sufficient [18] as it gives only 50 percent increase in the wall velocity, whereas we observe the increase in the velocity by two and a half orders of magnitude. The field stimulated motion is much slower. Besides, it follows the filed with large time delay, which is absent in case of current excitation. The low-temperature velocity of the leading edge of the domain wall is compatible with the spin-wave velocity, so we should suggest that the current flow completely removes the barriers along the way of the DW. The tailoring part of the wall moves with the velocity which is in agreement with theory predictions.

In conclusion, the high rate of the formation 90°-domain and its fast expansion, easy control of the domain expansion distance by the current magnitude and pulse duration, the domain stability can be used for developing a new type of sequentially switched serial device based on the hybrid superconductor-ferromagnet-superconductor structures. All one needs is to put the series of superconducting structures nearby the permalloy-niobium stripe so that superconductivity could be suppressed by the stray field coming from the expending 90°-domain, figure 6. The in-plane or out-of-plane components of the field could be used. The resistance in the structures would increase substantially from zero up to definite value providing infinite reversible variation of the resistance in the given number of elements.

## Acknowledgment

L.S.U. thanks Berdyugin A., Chugunov A., and Ponomareva A. for some technical assistance, and Guslienko K., Morozov A., Slavin A., Tikhomirov O.A, and Urazdin S. for useful discussions. The work was supported by Russian Foundation of Basic Research under Grant 15-02-06743 in part of improvement of technology of sample fabrication and by Russian Scientific Foundation under Grant 14-12-01290 in part of scientific research.

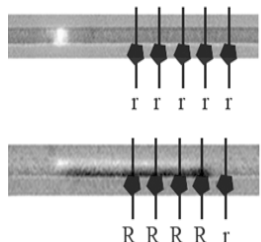

Fig. 6. Array of the elements (black circles) which could be swithed from the resistance *r* to resistance *R* by out-of-plane magnetic field produced by the shifted domain wall.